\documentclass{icrc2009}

\usepackage{graphicx}   % for including figures
\usepackage[font=footnotesize caption=false]{subfig} % subfig.sty for a double column floating figure using two subfigures
\usepackage{fixltx2e}
\usepackage{url}

\newcommand{\shorttitle}[1]%
{\markboth{Proceedings of the 31\MakeLowercase{$^{st}$} ICRC, {\L}\'{o}d\'{z} 2009}{#1} }
\newcommand{\etal}{\MakeLowercase{\textit{et al. }}} % "et al."

\newcommand{\gray}[1]{$\gamma$-ray{#1}}

\newcommand{\Xco}{$X_{\rm CO}$}

\newcommand{\fermi}{{\it Fermi}{}}

\begin{document}
\title{Fermi LAT Measurements of the Gamma-Ray Emission from the Large Magellanic Cloud}

\author{\IEEEauthorblockN{Troy A. Porter\IEEEauthorrefmark{1} and J\"{u}rgen Kn\"{o}dlseder\IEEEauthorrefmark{2} for the {\em Fermi} LAT Collaboration}
                            \\
\IEEEauthorblockA{\IEEEauthorrefmark{1}Santa Cruz Institute for Particle Physics, University of California, 1156 High Street, Santa Cruz 95064\\
\IEEEauthorrefmark{2}Centre d'\'{E}tude Spatiale des Rayonnements, CNRS/Universit\'{e} de Toulouse, PO Box 44346, 31028 Toulouse Cedex 4}}

% please write the preseter's name and short title (3-4 words maximum)
%    which will appear at the header of the even pages.
\shorttitle{Porter \etal LAT Measurements of LMC Gamma rays}
\maketitle

\begin{abstract}
Apart from the Milky Way, the Large Magellanic Cloud (LMC) is the only other 
normal star-forming galaxy that was conclusively detected in high 
energy ($> 100$ MeV) \gray{s} by the Energetic Gamma Ray 
Telescope (EGRET) on the {\em Compton Gamma-Ray Observatory}.
However, the sensitivity of EGRET was sufficient only to marginally resolve 
the LMC. 
We report on measurements of the \gray{} emission from the LMC by the Large 
Area Telescope (LAT) on the \fermi{} Gamma-ray Space Telescope. 
For the first time an externally viewed star-forming galaxy is well 
resolved in \gray{s}. 
We discuss the distribution of the LMC diffuse emission as seen by the 
LAT and implications for cosmic-ray (CR) physics.
\end{abstract}

 
\section{Introduction}

The diffuse Galactic emission (DGE) is the dominant feature of the \gray{} sky.
It is produced by interactions of 
CRs, mainly protons and electrons, 
with the interstellar gas (via $\pi^0$-production and Bremsstrahlung) and 
radiation field (via inverse Compton [IC] scattering).
It is a direct probe of CR fluxes in distant locations, and
may contain signatures of physics beyond the Standard Model, 
such as dark matter annihilation or decay.
However, the interpretation of the DGE is complicated by the fact that a large
number and variety of discrete sources of \gray{s} 
can also contribute, thus blurring the distinction between 
individual CR acceleration sites and true diffuse emission processes.

Gamma rays from CR interactions with the interstellar medium are also expected
from nearby galaxies, and indeed, the EGRET instrument on board the 
{\em Compton Gamma Ray Observatory} detected \gray{} emission from the LMC 
\cite{Sreekumar1992}.
The LMC is an excellent target for studying the link between CR acceleration
and propagation and diffuse \gray{} emission because of its 
proximity and its almost face-on orientation simplifies the interpretation
of the diffuse emission versus the DGE.
In addtion, the LMC is host to many supernova remnants, bubbles and 
superbubbles, and massive star-forming regions that all are potential CR 
acceleration sites \cite{Cesarsky1983,Biermann2004,Binns2007}.

The LAT onboard the \fermi{} Gamma-ray Space Telescope provides significantly 
improved capabilities for in-depth studies of 
the diffuse \gray{} emission from nearby galaxies, and of the LMC 
in particular \cite{Digel2000,Weidenspointner2007}.
In this paper we report on the initial analysis of measurements of the diffuse
\gray{} emission from the LMC taken in the course of the first year's 
all-sky survey by the LAT.

\section{Data Selection and Analysis}
The LAT is the primary instrument on \fermi{} which was launched from 
Cape Canaveral on June 11, 2008.
The LAT is an imaging, wide field-of-view, high-energy \gray{} telescope,
convering the energy range $\sim 20$ MeV to $> 300$ GeV, that operates 
according to the pair-conversion principle and is instrumented with a precision 
tracker and 
calorimeter, each consisting of a $4\times 4$ array of 16 modules, a segmented 
anti-coincidence detector (ACD) that covers the tracker array, 
and a programmable 
trigger and data acquisition system. 
The LAT has a large $\sim 2.4$ sr field of view, and compared to 
earlier \gray{} missions, has a large effective area ($> 7000$ cm$^2$ on axis 
at $\sim 1$ GeV for the event selection used in this paper), 
improved angular resolution ($\sim 0.5^\circ$ 
68\% containment radius at 1 GeV for \gray{} conversions in the 
thin-radiator section of the LAT) and low dead 
time ($\sim 25$ $\mu$s per event). 
The 1$\sigma$ energy resolution in the 100 MeV -- 10 GeV energy range is
better than $\sim 10$\%. 
Full details of the instrument, onboard and ground data processing 
are given in \cite{InstrumentPaper}.
%The on-orbit instrument calibration is presented in \cite{OnorbitCalibrationPaper}.

The data used in this work cover the period 
August 8th 2008 –- April 24th 2009 and amounts to 211.7 days of 
continuous sky survey data taking. 
During this period a total exposure 
of $\sim 2.3 \times 10^{10}$ cm$^2$ s (at 1 GeV) 
was accumulated for the LMC region.

\subsection{Data Preparation}

The analysis presented in this paper was performed using the 
LAT Science Tools package, that is available from the \fermi{} Science Support 
Centre, with post-launch instrument response functions (IRFs, designated P6V3).
%These take into account pile-up and accidental coincidence 
%effects in the detector 
%subsystems that are not considered in the definition of the pre-launch IRFs. 
%Cosmic rays, primarily protons,
%pass through the LAT at a high rate and sufficiently 
%near coincidences with \gray{s}
%leave residual signals that can result in \gray{s} being misclassified,
%particularly at energies $\leq 300$ MeV.  
%The post-launch IRFs were
%derived using LAT events read out at regular intervals 
%as a background overlay on
%the standard simulations of \gray{s} and provide an accurate
%accounting for the instrumental pile-up and accidental coincidence effects. 
Events for the data taking period mentioned above satisfying the 
standard low-background event selection (``Diffuse'' events 
\cite{PorterICRC0554,GeVExcessPaper}) and coming from 
zenith angles $< 105^\circ$ (to greatly reduce the contribution by 
Earth albedo
\gray{s}) were used.
To further reduce the effect of Earth albedo 
backgrounds, the time intervals when the Earth was appreciably within the 
field of view (specifically, when the centre of the field of view was 
more than 47$^\circ$ from the zenith) were excluded from this analysis.
%, and 
%all events taken when the spacecraft was within the South 
%Atlantic Anomaly were also excluded.
We further restrict 
our analysis to photon energies $> 200$ MeV where our current 
knowledge of the instrument response implies systematic uncertainties that 
are $< 10$\%.
% and where the photon energy dispersion due to incomplete 
%energy measurements becomes negligible.

\subsection{Morphology}

\begin{figure}[t]
  \centering
  \includegraphics[width=7.5cm]{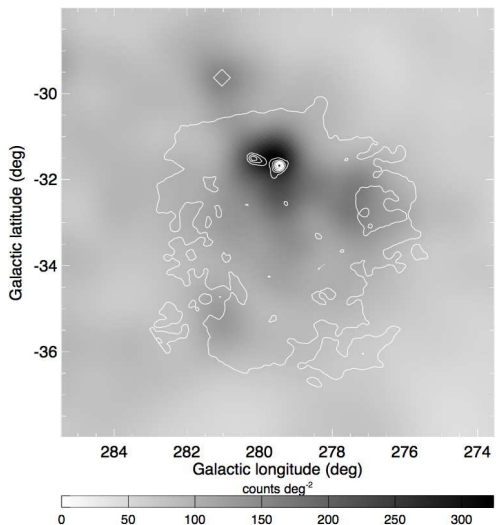}
  \caption{Preliminary adaptively smoothed (s.n.r. = 10) LAT counts map of a
    $10^\circ \times 10^\circ$ region centred on the LMC for the energy 
    range 200 MeV -- 100 GeV. Contours show the extinction map 
    from \cite{Schlegel1998} as an approximate 
    tracer of the total gas column density in the LMC. Ten linearly spaced 
    contour levels are plotted. 
    The diamond in the north-east of the image designates the 
    location of the blazar CRATES J060106-703606 \cite{Healey2007} that 
    contributes at a low level to the \gray{} emission in this area.
  }
  \label{fig1}
\end{figure}

Figure~\ref{fig1} shows a \gray{} counts map binned in $3\prime$ pixels 
over a $10^\circ \times 10^\circ$ area centred on the LMC.
The exposure is very uniform across the region shown.
The binned map has been smoothed using a two-dimensional
adaptive Gaussian kernel technique \cite{Ebeling2006} to reduce the effects
of Poissonian noise. 
The signal-to-noise ratio (s.n.r.) has been set to 10 to reduce
statistical noise variations below $< 10$\%.
Also shown is an overlay of the extinction map taken from 
\cite{Schlegel1998} (hereafter, SFD)
which is a tracer of the total gas column density in the LMC.
To first order, the extinction scales linearly with total gas column (note,  
the colour correction method used by \cite{Schlegel1998} breaks
down for the peak of the extinction map due to the massive HII region in
the centre of 30 Doradus).
A substantial fraction of the gas is in a small area to the north of the LMC 
at $(l,b) \approx (279.5^\circ, -31.5^\circ)$ which coincides with the 30 Doradus
star-forming region.
The LAT-measured LMC \gray{} emission also peaks in this area.
The \gray{} intensity in the 30 Doradus region 
exceeds $\sim 300$ counts deg$^{-2}$ while the intensity level in other regions
of the LMC is $\sim 120$ counts deg$^{-2}$ (the background rate around the 
LMC is $\sim 50 - 70$ counts deg$^{-2}$).

\begin{figure}[t]
  \centering
  \includegraphics[width=7.5cm]{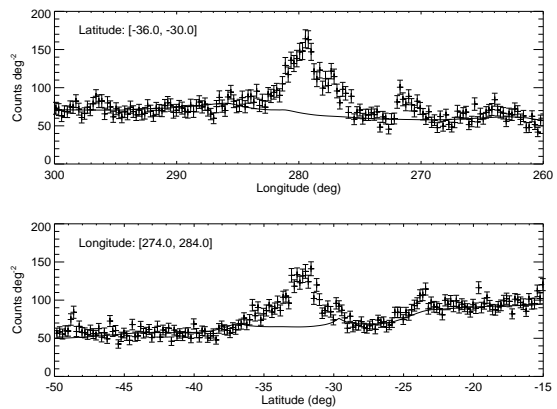}
  \caption{Preliminary 
    longitude (top) and latitude (bottom) photon intensity profiles of
    the LMC region for the energy range 200 MeV -- 100 GeV. 
    The solid line indicates the expected contributions from DGE, 
    diffuse extragalactic emission, instrumental background and the 
    blazar CRATES J060106-703606 at $(l, b) = (281.04^\circ, −29.63^\circ)$ 
    in this area of the sky.
  }
  \label{fig2}
\end{figure}

The excess in the 30 Doradus region is also seen in the longitude and latitude
profiles of the LAT-measured \gray{} intensity shown in Fig.~\ref{fig2}.
Within the rectangular region covering $274^\circ \leq l \leq 284^\circ$ and 
$-36^\circ \leq b \leq -30^\circ$ the net counts are $\sim 1800$ 
in the energy range 200 MeV -- 100 GeV above the expected contributions
from DGE, extragalactic diffuse emission, residual charge-particle 
background, and the 
blazar CRATES J060106-703606 that was active for part of the data taking 
period.
These background contributions have been estimated by fitting the data with 
spatial and 
spectral templates of the different components together with a spatial template
for the LMC emission using a maximum likelihood method.
The DGE has been modelled using the GALPROP code \cite{SMR04,Porter2008} 
with a model corresponding
to the GALDEF file 54\_59Xvarh7S\footnote{Available from the 
website: http://galprop.stanford.edu}.
For the combination of extragalactic diffuse emission and residual 
background we assumed an isotropic component with a power-law spectrum.
CRATES J060106-703606 is modelled as a point source at
$(l, b) = (281.04^\circ , −29.63^\circ)$ with a power-law spectrum. 
For the LMC, we used the SFD extinction map as the 
spatial template.
The average magnitude of the extinction map is obtained by averaging over all
pixels outside a radius of $4^\circ$ 
around $(l, b) = (279.65^\circ , −33.34^\circ)$ within the rectangular region.
For this figure, the pedestal value was subtracted from all pixels in the map.
Then, all pixels outside of a radius of $4^\circ$ 
centred on $(l, b) = (279.65^\circ , −33.34^\circ)$ are set to zero in
order to extract the LMC emission. 
For the LMC spectral model we assumed a power law.

To describe the morphology of the \gray{} emission from the LMC we initially
fit a point source with the position and flux free on top of the background
model\footnote{We call the combination of the GALPROP model, the 
isotropic model, and the CRATES J060106-703606 point source 
the ``background model'' for our analysis. The free parameters of this 
background model are the normalisation of the GALPROP model, the intensity 
and spectral slope of the isotropic component, and the flux and 
spectral slope of the CRATES J060106-703606 point source.} to the data.
This results in a best-fit point-source position 
$(l,b) = (279.58^\circ, -31.72^\circ)$ with an error radius $0.09^\circ$ 
95\% confidence (statistical; the systematic position uncertainty is estimated
to be $< 0.02^\circ$).
This position is close to the central star cluster of 30 Doradus, R136, located
at $(l,b) = (279.47^\circ, -31.67^\circ)$ which is $\sim 0.11^\circ$ from our
best-fit point-source location.

The detection significance of the LMC can be estimated using the so-called
Test Statistic ($TS$) which is defined as twice the difference 
between the log-likelihood, $\cal{L} _{\rm 1}$, that is obtained from fitting the
LMC model on top of the background model to the data, and the 
log-likelihood,$\cal{L} _{\rm 0}$ that is obtained by fitting the
background model only, $TS = 2(\cal{L}_{\rm 1} - \cal{L}_{\rm 0})$. 
Under the hypothesis that our model satisfactorily explains the LAT data, 
TS follows a $\chi _p ^2$-distribution
with $p$ degrees of freedom, where $p$ is the number of free 
parameters in the LMC model \cite{Cash1979}. 
For the case of a point source with position, flux, and spectral index free 
we obtain $TS = 869.1$.

We replaced the point source with an extended source model which was 
implemented as an axisymmetric two-dimensional Gaussian with variable angular
size $\theta$.
We redid the fit for position, flux, spectral index, and angular size and 
obtained a best-fitting source position $(l,b) = (279.5^\circ, -32.3^\circ)$ 
(95\% confidence radius $0.1^\circ$) and source extent 
of $\theta = 1.0^\circ \pm 0.1^\circ$.
The $TS$ value is 1088.5 with a corresponding significance for source 
extension compared with the point source hypothesis of $14.8\sigma$ 
($p = 5 - 4 = 1$).

\begin{table}[!h]
  \caption{Comparison of maximum likelihood model fitting results}
  \label{tablemaps}
  \centering
  \begin{tabular}{|c|c|c|}
    \hline
    LMC Model  & $TS$ & Number of Parameters \\
    \hline 
    Point source & 869.1 & 4 \\
    2D Gaussian source & 1088.5 & 5 \\
    HI gas map & 1173.4 & 2 \\
    CO gas map & 932.2 & 2 \\
    HI + CO gas map & 1176.1 & 4 \\
    SFD extinction map & 1179.6 & 2 \\
    IRIS 100 $\mu$m map & 1179.1 & 2 \\
    \hline
  \end{tabular}
\end{table}

In addition to the geometrical models we compared the LAT data to 
various spatial templates that trace the interstellar matter distribution 
in the LMC.
For HI we use the aperture synthesis and multibeam data that \cite{Kim2005}
have combined from ATCA and Parkes observations.
For H$_2$ we use the CO observations of the LMC obtained with the NANTEN
telescope \cite{Yamaguchi2001}.
We also used the SFD extinction map as tracer of the total
gas column density and the 100 $\mu$m IRIS map that
has been obtained by reprocessing the IRAS survey data \cite{irispaper}.
The results of this comparison are summarised together with
that of the geometrical models in Table~\ref{tablemaps}.
The best fits are obtained for the SFD extinction map and the IRIS 100 $\mu$m
map which give $TS$ values of 1179.6 and 1179.1, respectively. 
For 2 free parameters (the total flux in the map and the spectral index) 
this corresponds to a detection significance of $34.5\sigma$. 
An almost equally good fit is obtained using the HI map. 
Fitting the CO map to the LAT data provides a poor fit, 
suggesting that the \gray{} morphology differs
from that of molecular gas in the LMC. 
Fitting the HI and CO maps together
confirms this result since $\sim 97$\% 
of the total flux is attributed to
the HI component.
The corresponding increase in $TS$ with respect to simply fitting the HI map 
is negligible.

The HI/SFD/IRIS 100 $\mu$m maps fit the data considerably better than
a single point source, adding further evidence that the observed 
high-energy \gray{} emission is extended. 
Furthermore, the two-dimensional Gaussian source model is not as good as 
the tracer maps, suggesting that the emission morphology is more complex 
than a single Gaussian shape.

\subsection{Spectrum}

\begin{figure}[t]
  \centering
  \includegraphics[width=7.5cm]{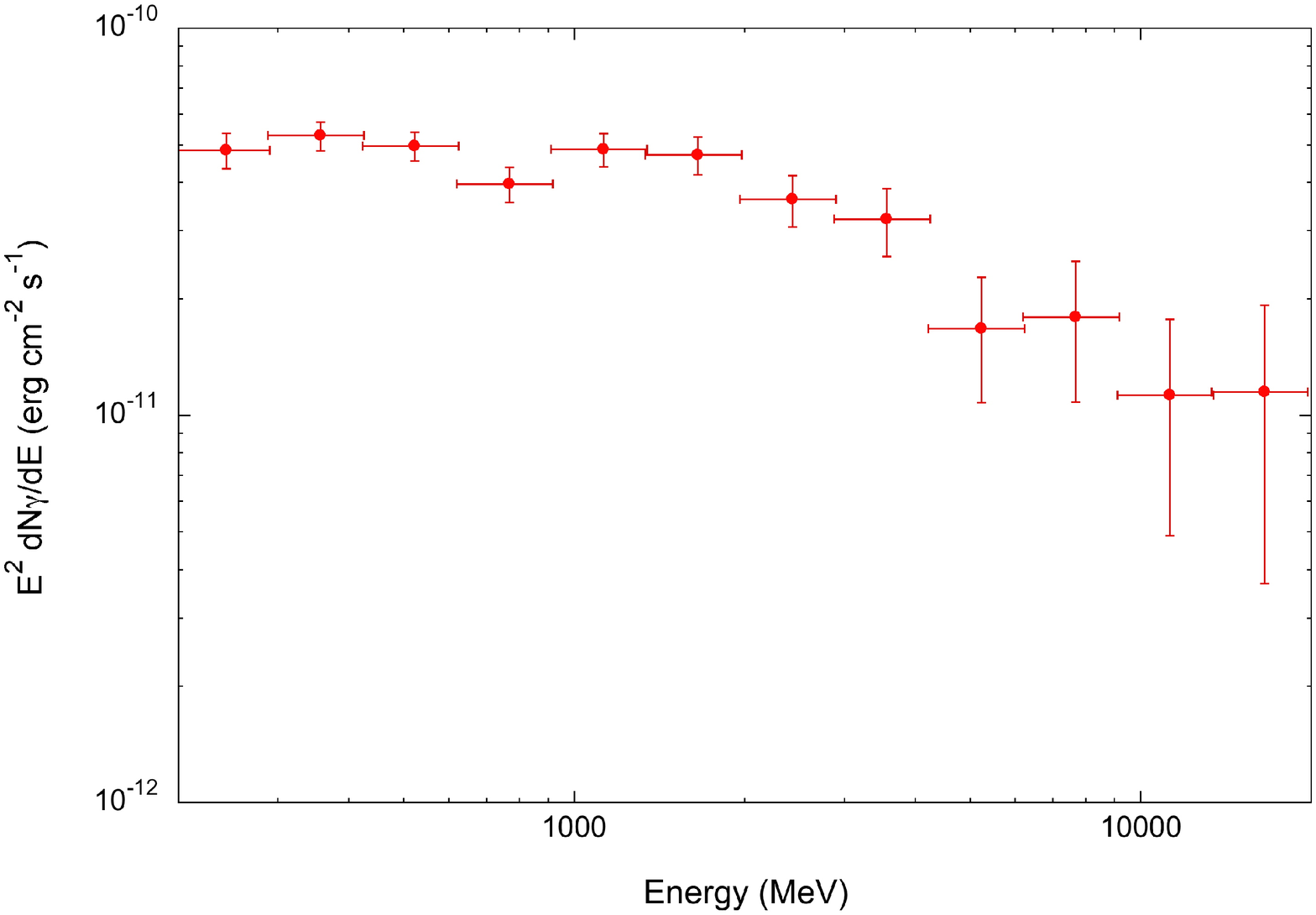}
  \caption{Preliminary LMC spectrum obtained by fitting the extinction map 
    from \cite{Schlegel1998} in 12 logarithmically-spaced energy bins 
    covering the
    energy range 200 MeV -- 20 GeV to the LAT data. 
    Note that errors are statistical only.
  }
  \label{fig3}
\end{figure}

Using the SFD extinction map (i.e., our best fitting 
spatial template of the high-energy emission) 
we extract a spectrum for the LMC by
fitting the data in 12 logarithmically-spaced energy bins covering the 
energy range 200 MeV -- 20 GeV. 
Above 20 GeV, photons from the LMC become too sparse in the data 
set used in this paper to allow for meaningful spectral points to be derived.

Figure~\ref{fig3} shows the LMC spectrum obtained by this method.
Our analysis indicates a spectral steepening of the emission with 
increasing energy, suggesting a simple power law is an inadequate 
description of the data. 
We confirm this trend by fitting the data using a broken power law 
instead of a simple power law. 
This results in an improvement of the $TS$ by 10.1 with respect to a simple
power law,
corresponding to a significance of $2.7\sigma$ ($p = 2$) for spectral 
steepening. 
Alternatively, fitting an exponentially cutoff power law 
improves the $TS$ by 7.8 with respect to a simple power law, corresponding 
to a significance of $2.8\sigma$ ($p = 1$) for a spectral cutoff.
Integrating either model over the energy range 100 MeV -- 500 GeV gives 
identical photon fluxes of $(3.1 \pm 0.2) \times 10^{-7}$ 
ph cm$^{-2}$ s$^{-1}$ and 
energy fluxes of $(2.0 \pm 0.1) \times 10^{-10}$ erg cm$^{-2}$ s$^{-1}$ for 
the LMC.
The systematic uncertainty in these flux measurements is $\sim 10$\%.

\section{Discussion}

The EGRET team \cite{Sreekumar1992} reported the initial detection of the 
LMC in high-energy \gray{s} based on 4 weeks of data.
Due to the limited angular resolution and sensitivity, the emission detected
from the LMC was weak and details of the spatial structure of the
galaxy were not resolved.
However, it was obvious from the EGRET data that the LMC was an extended
\gray{} source.

The LAT now clearly resolves the \gray{} emission from the LMC.
The maximum of the emission is attributable to the massive star-forming 
region 30 Doradus.
While this coincidence could be taken as a hint for an enhanced
CR density in 30 Doradus with respect to the rest of the LMC, we note
that a substantial fraction of the interstellar gas of the LMC is also 
confined to the 30 Doradus area.
Consequently, the target density for CR interactions
is greatly enhanced in this region which implies a corresponding enhancement
of the \gray{} luminosity. 
In addition, the radiation field in this region is more intense due to the 
presence of significant numbers of massive, young stars, and this could 
also result in enhanced IC emission. 
Whether the data also support an enhanced CR density in 30 Doradus with 
respect to the rest of the galaxy requires further analysis.

The poor fit of the CO map to the LAT data 
suggests that the overall
distribution of \gray{} emission differs from that of the H$_2$ in the LMC.
The distribution of HI fits the data considerably better and the
combined fit of HI and CO maps indicates that any contribution to the \gray{} 
emission that is correlated to the molecular gas is at best marginal. 
This agrees well with expectations since the gas budget of the LMC is 
largely (90-95\%) dominated by HI \cite{Fukui1999}.
Consequently we are
presently unable to determine the CO-to-H2 conversion factor, \Xco, from our
LMC data.

Based on the assumptions of dynamic balance and that the electron/proton 
ratio in the LMC is the same as in the Milky Way \cite{Fichtel1991}
predicted a \gray{} flux $> 100$ MeV for the LMC galaxy of 
$F_{\rm LMC}(> 100 \, {\rm MeV}) = 2.3 \times 10^{-7}$ ph cm$^{-2}$ s$^{-1}$.
More recently, \cite{Pavlidou2001} predicted a \gray{} flux $> 100$ MeV 
$F_{\rm LMC}(> 100 \, {\rm MeV}) = 1.1 \times 10^{-7}$ ph cm$^{-2}$ s$^{-1}$ based
on estimates of the LMC supernova rate and total gas densities.
The flux measured by the LAT of $(3.1 \pm 0.2) \times 10^{-7}$ ph 
cm$^{-2}$ s$^{-1}$ is greater than these
model estimates.
However, given the uncertainties in the models the agreement is satisfactory.
     
Further studies of the LMC with the LAT will concentrate on the
spectral analysis of the data, placing particular emphasis on variations of 
the spectrum throughout the LMC. 
Due to the excellent sensitivity and angular resolution of the LAT this is 
the first time that such studies are possible. 
In addition, other nearby galaxies await detection as \gray{} sources, 
such as the Small Magellanic Cloud and the Andromeda Galaxy (M31). 
Both should be detectable by the LAT and the comparative study of 
their diffuse \gray{} emission will further our knowledge of CR physics in 
nearby galactic environments and ultimately our own Galaxy.

{\it Acknowledgements:}
The \fermi{} LAT Collaboration acknowledges support from a number of 
agencies and institutes for both development and the operation of the LAT 
as well as scientific data analysis. 
These include NASA and DOE in the United States, CEA/Irfu and IN2P3/CNRS 
in France, ASI and INFN in Italy, MEXT, KEK, and JAXA in Japan, and 
the K.~A.~Wallenberg Foundation, the Swedish Research Council and the 
National Space Board in Sweden. 
Additional support from INAF in Italy for science analysis during the 
operations phase is also gratefully acknowledged.

\end{document}